# Interplay between Intra- and Intermolecular Charge Transfer in the Optical Excitations of J-Aggregates

Michele Guerrini,[†,‡,§] Caterina Cocchi,*,§ Arrigo Calzolari,[‡] Daniele Varsano,*,‡ and Stefano Corni[‡,∥]

†Dipartimento FIM, Università di Modena e Reggio Emilia, 41125 Modena, Italy
‡CNR Nano Istituto Nanoscienze, Centro S3, 41125 Modena, Italy
§Department of Physics and IRIS Adlershof, Humboldt-Universität zu Berlin, 12489 Berlin, Germany
∥Dipartimento di Scienze Chimiche, Università di Padova, 35131 Padova, Italy

Ⓢ Supporting Information

**ABSTRACT:** In a first-principles study based on density functional theory and many-body perturbation theory, we address the interplay between intra- and intermolecular interactions in a J-aggregate formed by push–pull organic dyes by investigating its electronic and optical properties. We find that the most intense excitation dominating the spectral onset of the aggregate, i.e., the J-band, exhibits a combination of intramolecular charge transfer, coming from the push–pull character of the constituting dyes, and intermolecular charge transfer, due to the dense molecular packing. We also show the presence of a pure intermolecular charge-transfer excitation within the J-band, which is expected to play a relevant role in the emission properties of the J-aggregate. Our results shed light on the microscopic character of optical excitations of J-aggregates and offer new perspectives to further understand the nature of collective excitations in organic semiconductors.

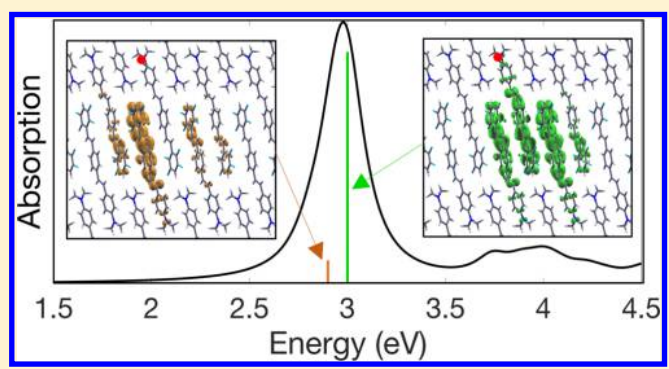

## ■ INTRODUCTION

J-aggregates are a special class of molecular crystals with enhanced light−matter interaction properties.[1−5] Their optical spectra are dominated by a very strong and narrow peak at the onset—the so-called J-band—which appears at lower energy with respect to the isolated molecular constituents.[6−10] This peculiar feature emerges as a collective effect of the monomers in the aggregated phase and is typically explained in terms of coherent intermolecular dipole coupling.[6,8,9,11,12] The microscopic nature and the fundamental mechanisms that give rise to the J-band are still debated.[6] The situation is even more complex in the case of J-aggregates formed by polar molecules like push−pull organic dyes. In this case, intramolecular charge-transfer (CT) adds up to the aforementioned intermolecular interactions that are responsible for the formation of the J-band. An example of this kind of system is the J-aggregate formed by the organic chromophore 4-(N,N-dimethyl-amino)-4-(2,3,5,6-tetra-fluorostyryl)-stilbene ($C_{24}H_{19}F_4N$), which has been recently proposed and synthesized by Botta et al.[13] Some of the authors have recently investigated this J-aggregate by means of time-dependent density functional theory (TDDFT) showing that its optical behavior cannot be deduced by considering only its isolated components due to the intrinsically supramolecular nature of its optical response.[14] On the other hand, the character of the excitations forming the J-band and the interplay between inter- and intramolecular interactions still need to be addressed. This issue is relevant in the broader context of the electronic structure characterization of molecular crystals. Even in the case of nonpolar molecules, the discussion regarding the nature of optical excitations in organic semiconductors[15−19] is still ongoing. Both localized Frenkel excitons and delocalized intermolecular excitations can appear at the spectral onset of organic semiconductors: Their relative energetic position has been rationalized in terms of intermolecular interactions and wave-function overlap.[19] Identifying the character of the lowest-energy excitations in molecular crystals is particularly relevant to interpret phenomena like multiple exciton generation and singlet fission that have been recently observed in these systems[20−32] and that promise a breakthrough in view of optoelectronic applications. Addressing this question from a theoretical perspective requires a high-level methodology that incorporates a reliable description of the electronic structure and excitations including electron−electron and electron−hole (e−h) correlation effects.

Many-body perturbation theory (MBPT), based on the GW approximation and the solution of the Bethe−Salpeter equation (BSE), is the state-of-the-art approach to investigate optical excitations in crystalline materials. In the last two decades, it has been successfully applied also to molecular systems, providing unprecedented understanding on the character of the excitations therein.[15−19,33−39] Many of these studies are focused on oligoacenes,[15,16,19,35] which have drawn particular attention since the last years of the past century for their appealing optoelectronic and transport properties.[40−47] The first-principles works cited above have revealed that the low-energy excitations in these materials exhibit a remarkable excitonic character, with binding energies of the order of a few hundreds of milli-electron volts and an intermixed Frenkel-like and intermolecular charge-transfer character. Intermolecular interactions generally play a decisive role in the optical properties of a variety of molecular aggregates.[48,49] For example, in azobenzene-functionalized self-assembled monolayers, intermolecular interactions impact strongly on light absorption and excitonic coupling, and hence critically influence the photoisomerization process.[50−53] In these systems, intermolecular coupling and packing effects have been shown to be crucial also in core excitations.[54]

In this work, we adopt the formalism of MBPT to investigate the nature of the excitations in a J-aggregate formed by C$_{24}$H$_{19}$F$_4$N dyes (Figure 1), devoting specific attention to the interplay between intra- and intermolecular charge transfer. MBPT calculations are carried out on top of density functional theory (DFT) on the J-aggregate and, for comparison, also on its isolated molecular unit. With the analysis of the quasi-particle (QP) electronic structure and the optical absorption spectra including excitonic effects, we characterize the excited states of this J-aggregate with unprecedented insight. Specifically, we focus on two excited states with different characters appearing within the J-band: the first one gives rise to the main peak of the J-band and manifests a mixed intra- and intermolecular charge-transfer nature; the second one is a very weak excited state possessing a dominant intermolecular CT behavior. This analysis demonstrates how the interplay between intra and intermolecular charge transfer determines the optical properties in a molecular J-aggregate and contributes to the more general understanding of the nature of collective excitations in molecular crystals.

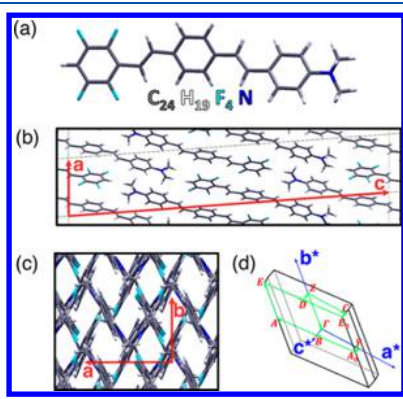

**Figure 1.** (a) Push−pull organic dye 4-(N,N-dimethyl-amino)-4-(2,3,5,6-tetra-fluorostyryl)-stilbene (C$_{24}$H$_{19}$F$_4$N) and its J-aggregate viewed (b) from the a−c plane and (c) from the a−b plane, with the lattice vectors marked in red; crystallographic structure from CCDC no. 961738 and ref 13. (d) Brillouin zone (BZ) associated with the unit cell of the J-aggregate with the reciprocal lattice vectors indicated in blue, the high-symmetry points highlighted in red, and the path chosen for the band structure plot marked in green.

## ■ METHODS

**Theoretical Background.** This study is based on DFT[55,56] and MBPT (the GW approximation and the Bethe−Salpeter equation).[57−59] The workflow adopted to calculate the electronic and optical properties proceeds through three steps: First, a DFT calculation is performed to compute Kohn−Sham (KS) single-particle energies and wave functions as a basis in the successive steps; next, the quasi-particle correction to the KS energies is calculated through a single-shot GW calculation;[57,58,60] the Bethe−Salpeter equation is solved to obtain optical absorption spectra together with excitonic eigenfunctions and eigenenergies.[57] This approach ensures state-of-the-art accuracy methods in the calculation of excited-state properties and, concomitantly, a quantitative insight into the character of the excitations.

In solving the BSE, we adopt the Tamm−Dancoff approximation (TDA)[57] consisting in diagonalizing an effective excitonic Hamiltonian

$$\hat{H}_{ex} = \sum_{ck} \epsilon_{ck} a_{ck}^\dagger a_{ck} - \sum_{\upsilon k} \epsilon_{\upsilon k} b_{\upsilon k}^\dagger b_{\upsilon k}$$
$$+ \sum_{\substack{\upsilon c k \\ \upsilon' c' k'}} (2\bar{\upsilon}_{\upsilon' c' k'}^{\upsilon c k} - W_{\upsilon' c' k'}^{\upsilon c k}) a_{ck}^\dagger b_{\upsilon k}^\dagger b_{\upsilon' k'} a_{c' k'}$$

(1)

where the index $\upsilon(c)$ indicates valence (conduction) bands, $\mathbf{k}$ and $\mathbf{k}'$ are wave vectors in the Brillouin zone (BZ), and $a^\dagger(a)$ and $b^\dagger(b)$ are creation (annihilation) operators for electrons and holes, respectively. The quasi-particle energies $\epsilon_{ck}$ and $\epsilon_{\upsilon k}$ are obtained from the GW calculation. The matrix elements of $\bar{\upsilon}$, which is the short-range unscreened Coulomb interaction, and of $W$, the statically screened Coulomb interaction, represent the exchange and direct part of the BSE Hamiltonian of eq 1 and are expressed as $\bar{\upsilon}_{\upsilon' c' k'}^{\upsilon c k} = \langle ck, \upsilon' k' | \bar{\upsilon} | \upsilon k, c' k' \rangle$ and $W_{\upsilon' c' k'}^{\upsilon c k} = \langle ck, \upsilon' k' | W | c' k', \upsilon k \rangle$, respectively. The latter takes into account the screened electron−hole interaction (i.e., excitonic effects), while the former is responsible for crystal local-field effects (the factor 2 derives from spin summation in the singlet channel). Note that the potential $\bar{\upsilon}$ is a modified Coulomb interaction defined as the bare Coulomb interaction without the long-range (i.e., macroscopic) contribution (i.e., $\lim_{q\to0} \bar{\upsilon}_{G=0}(q) = 0$ in reciprocal space).[61] To quantify local field effects (LFEs) for a certain excited eigenstate $|\lambda\rangle$ of the excitonic Hamiltonian (eq 1) with excitation energy $E^\lambda$, we define the e−h exchange energy $E_x^\lambda = \sum_{\substack{\upsilon c k \\ \upsilon' c' k'}} A_{\upsilon c k}^\lambda (A_{\upsilon' c' k'}^\lambda)^* \bar{\upsilon}_{\upsilon' c' k'}^{\upsilon c k}$, where $A_{\upsilon c k}^\lambda$ are the BSE coefficients of the eigenstate $|\lambda\rangle$). When the total excitation energy $E^\lambda$ is below the QP gap, i.e., $E^\lambda - E_{GW}^{gap} < 0$, the excited state $|\lambda\rangle$ is considered a bound exciton with a binding energy defined as $E_b = E_{GW}^{gap} - E^\lambda$, which physically represents the energy required to separate a bound electron−hole pair into a free electron and a free hole.

**Computational Details.** The unit cell of the bulk J-aggregate crystal, composed of 192 atoms, corresponding to 4 nonequivalent push−pull molecules, has been taken from the experimental X-ray structure available in the CCDC no. 961738 (for more details, see also the Supporting Information (SI) of ref 13) without any further relaxation. As previously shown in the case of the pentacene crystal,[16] the optical properties of organic semiconductors are rather insensitive to the optimization of the internal geometry. Moreover, the experimental X-ray structure already takes into account all of the supramolecular effects (e.g., van der Waals, π-stacking, hydrogen bonds, etc.), which determine the relative orientation and the packing of the molecules, is a reliable starting point for our calculations.

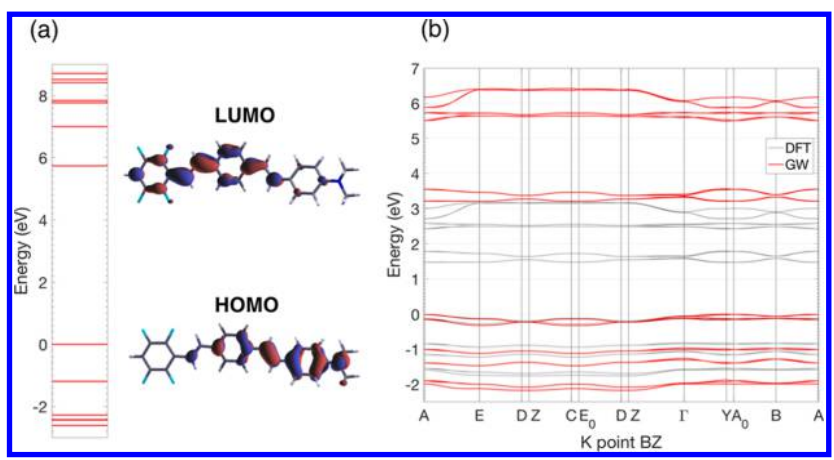

**Figure 2.** (a) GW energy levels (HOMO level set to zero) and frontier molecular orbitals of the isolated push–pull dye. The isovalues are fixed at 0.04 bohr$^{-3/2}$. (b) Band structure of the J-aggregate computed from DFT (gray) and one-shot GW (red), where in the latter, a scissors plus stretching has been applied to the DFT energy levels. The Fermi energy is set to zero at the GW valence band maximum (VBM).

DFT calculations of the J-aggregate were performed with the plane-wave code Quantum Espresso[62] by using the semilocal Perdew–Burke–Ernzerhof (PBE)[63] exchange–correlation (xc) functional and treating core electrons with a norm conserving pseudopotential.[64] The plane-wave cutoff for the Kohn–Sham (KS) wave functions (density) was fixed to 40 Ry (160 Ry), and the convergence criterion on the total energy was fixed to 10$^{-8}$ Ry. The mean residual force per atom is 0.01 Ry/bohr due to nonequilibrium C–H bond lengths, which are however not expected to alter the electronic structure of the system and the essence of our results. GW and BSE calculations were performed using the plane-wave code Yambo.[65] QP corrections were calculated by single-shot GW (i.e., perturbative $G_0W_0$) and adopting the Godby–Needs plasmon pole approximation (GN-PPA) model[66] to approximate the frequency dependency of the inverse dielectric function. The GN-PPA is considered a reliable approach for different kinds of bulk systems for which it yields results comparable to the more computationally demanding contour deformation approach.[67–69] Since the QP corrections obtained from $G_0W_0$ were almost constant for all of the KS states around the gap, in building the BSE Hamiltonian (eq 1), we applied a scissors plus stretching correction for all of the energies within the irreducible BZ by linear fitting of the QP-corrected values from $G_0W_0$ (scissors operator of 1.73 eV and stretching factors $S_v$ = 1.258 and $S_c$ = 1.216 for the occupied and unoccupied states, respectively). The QP gap of the J-aggregate obtained from $G_0W_0$ was checked against the one obtained with eigenvalue-only self-consistent GW calculations. The difference between the band gaps obtained with these two methods turned out to be in the order of 160 meV (see Section 2 of the Supporting Information (SI)). The BSE Hamiltonian of the J-aggregate was evaluated and diagonalized within the Tamm–Dancoff approximation using an e–h basis composed of 27 occupied and 34 unoccupied states and a $5 \times 3 \times 2$ $k$-point grid to sample the BZ. The final absorption spectrum of the bulk crystal was obtained by averaging the spectra computed for three orthogonal electric field polarizations (i.e., along $a$, $b$ axes and along the component of the $c$ axis perpendicular to the $a$–$b$ plane in Figure 1).

To treat the gas-phase push–pull molecule at the same level of theory as the J-aggregate, we computed the electronic and optical properties for the monomer as extracted from the J-aggregate without any further relaxation. In Section 1 of the SI, we show for the isolated gas-phase monomer that the main effect of the minimization of interatomic forces is to blue-shift the main absorption peak by a few hundreds of milli-electron volts without changing its character. To investigate the electronic and optical properties of the gas-phase push–pull dye, we used MOLGW,[70] a code that implements DFT and MBPT using localized basis orbitals, which are computationally more suited and efficient for isolated systems compared to plane waves. These DFT calculations are carried out using the PBE functional, cc-PVTZ basis set, and a full-electron treatment with frozen-core approximation and total energy convergence precision fixed to 10$^{-8}$ Ry. The QP correction was computed by an eigenvalue-only self-consistent GW calculation to minimize the dependence on the approximation for the xc functional used in the DFT starting point.[71] To calculate the correlation part of the self-energy, the MOLGW code adopts an analytical expression exploiting the spectral decomposition of the screened Coulomb potential and the residue theorem.[72] The following BSE step was solved in the TDA over a transition space of 50 occupied and 100 unoccupied orbitals. Additional information about the effects of different approximations and codes adopted in these calculations is reported in the SI.

## ■ RESULTS AND DISCUSSION

We start our analysis by inspecting the electronic structure of the isolated push–pull dye (Figure 1a) and of its J-aggregate (Figure 1b–d). In Figure 2a, the quasi-particle energy levels of the isolated organic dye are shown, together with the isosurface plots of the frontier molecular orbitals (MOs). The dimethylamino (push) and fluorinated ring (pull) groups at the opposite ends of the π-conjugated chromophore are responsible for the polarization on the frontier orbitals and the intrinsic dipole. The highest occupied molecular orbital (HOMO) and the lowest unoccupied molecular orbital (LUMO) are mostly localized on the electron-donating (push) and electron-withdrawing (pull) sides of the molecule, respectively. The QP band structure shown in Figure 2b exhibits a direct gap of 3.2 eV at the Y point, which is more than twice as large as the corresponding DFT (PBE) value of 1.48 eV and is half the QP gap of the isolated molecule, which is 5.66 eV in Figure 2a. The QP correction manifests itself as an almost rigid shift of 1.7 eV for all the conduction bands. This behavior is consistent with that of other organic crystals like pentacene and polythiophene.[18,73] As expected for molecular crystals,[15,16,39,74] the band dispersion is very limited and quite anisotropic: bands are almost flat along those directions in which intermolecular interactions are negligible and the wave-function overlap is hence minimized.

Conversely, along the $\pi$-stacking directions (e.g., along the Z−Γ−Y path in the BZ), intermolecular interactions are enhanced. As a result, bands are slightly more dispersive and band splitting is observed. Due to the presence of four inequivalent molecules in the unit cell, the bands close to the gap, which exhibit $\pi-\pi^*$ character, are almost degenerate. In Figure 3, the square moduli of the wave functions at the valence band maximum (VBM) and at the conduction band minimum (CBM) at the high-symmetry point Y are reported. The localized character of the frontier MOs found in the isolated push−pull dye is preserved also in the J-aggregate: The KS states corresponding to the VBM and CBM are mostly localized on the push and pull ends of each monomer, respectively. Figure 4 displays the optical absorption spectra of the isolated push−pull dye in the gas phase (Figure 3a) and of its J-aggregate (Figure 3b). In these plots, the red bars mark the analyzed excited-states. In Tables 1 and 2, we report information about the analyzed excited states for the monomer and the J-aggregate, respectively. The red shift of the principal peak of the J-aggregate with respect to the one of the single molecule amounts to 0.39 eV. This result is in better agreement with the experimental value of 0.48 eV,[13] compared to the outcome of TDDFT calculations with a semilocal xc-functional[14] (0.11 eV). The BSE results reveal also a rich excitonic structure underneath the J-band, which turns out to be composed of several transitions not shown in Figure 4b. Most of these excitations appear below the QP gap at 3.2 eV, and their binding energies do not exceed 0.4 eV, in line with the values reported for conjugated polymers[17,75–78] and slightly smaller than those of crystalline pentacene (0.5 eV),[16,19] picene (0.7 eV),[19] and antracene (0.8 eV).[15]

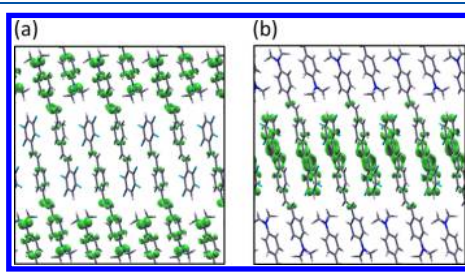

**Figure 3.** Isosurfaces of the squared modulus of the KS wave functions of the (a) valence band maximum and (b) conduction band minimum of the J-aggregate computed at the high-symmetry point Y. Isovalues fixed at 0.001 bohr$^{-3}$.

In the following, we focus on two representative excited states within the J-band: the first one labeled as J, at 3 eV, corresponding to the J-band peak; the second one labeled as $J_{CT}$, at 2.89 eV, which has a very low oscillator strength and represents a paradigmatic example of several intermolecular charge-transfer excited states within the J-band. Both excitations are analyzed in Figure 5 in terms of their excitonic wave functions and transition densities. Information about additional excited states of interest for the J-aggregate is reported in Section 2 of the SI. In Table 2, the composition of the J excitation is reported, showing that it is mainly formed by transitions between (quasi-)degenerate bands at the valence band top and conduction band bottom, which carry $\pi$ and $\pi^*$ characters like the HOMO and LUMO of the isolated dye (see Figures 2a and 3). The excitonic probability density of the excitation J (Figure 5a,b) reveals its intermixed intramolecular charge transfer as well as intermolecular charge transfer between nearest-neighboring molecules. Having fixed the position of the hole on the push side (i.e., the dimethylamino group) of one molecule in the unit cell, the electron is found on the pull part (i.e., the fluorinated ring) of the same molecule and of its nearest neighbors. By inspecting Figure 5a,b, the electron distribution is also delocalized around neighboring molecules along the $\pi$-stacking direction (i.e., the $a−b$ plane), where the dispersion is more pronounced because of the enhanced wave-function overlap. The intermolecular character of this excitation is therefore due to the $\pi$-interactions in the molecular packing direction, in analogy with the excitons of

   

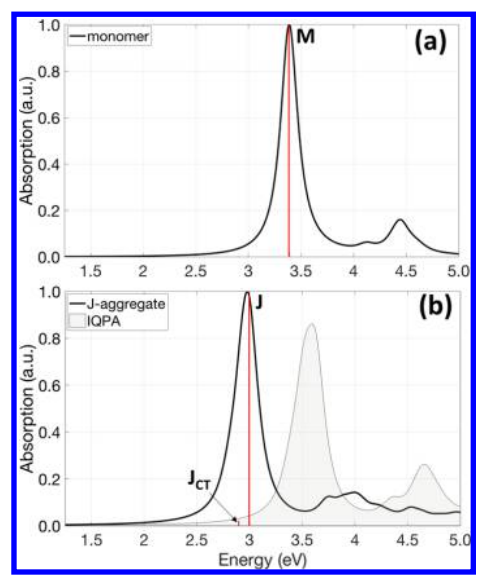

**Figure 4.** Absorption spectra of (a) the push−pull dye in the gas phase and (b) in the J-aggregate. The excited states analyzed in the text (M, J, $J_{CT}$) are marked by red bars with height representative of the relative oscillator strengths. The absorption in both BSE spectra is normalized with respect to the maximum value in the examined energy window. In (b), the absorption spectrum calculated from the independent quasi-particle approximation is also shown (gray-shaded area). All absorption spectra include a Lorentzian line width of 200 meV.

**Table 1. Energy and Composition of the First Excited State of the Push−Pull Monomer $C_{24}H_{19}F_4N$, Including the Associated Transition Energy and Weight Given by the Square of the BSE Coefficient**

| excited state energy (eV) | occupied level | unoccupied level | $\epsilon_c^{QP} - \epsilon_v^{QP}$ (eV) | weight $\|A_{cv}^A\|^2$ |
|---|---|---|---|---|
| $E^M = 3.39$ | HOMO | LUMO | 5.73 | 0.83 |

**Table 2. Energy and Composition of the J and $J_{CT}$ Excitations of the J-Aggregate, Including the Associated Transition Energy, the Associated k-Point in the Brillouin Zone Expressed in Reciprocal Crystal Units (RCUs), and the Weight Given by the Square of the BSE Coefficient[a]**

| excited state energy (eV) | occupied band | unoccupied band | k-point (RCU) | $\epsilon_{ck}^{QP} - \epsilon_{vk}^{QP}$ (eV) | weight $|A_{cvk}^{\lambda}|^2 \times 10$ |
|---|---|---|---|---|---|
| $E^J = 3$ | VBM − 2 | CBM + 1 | (0.4, 0, −0.5) | 3.34 | 0.59 |
| | VBM − 2 | CBM + 1 | (−0.4, 0, 0.5) | 3.34 | 0.56 |
| | VBM − 3 | CBM | (0.4, 0, 0) | 3.35 | 0.54 |
| | VBM − 3 | CBM | (−0.4, 0, 0) | 3.35 | 0.52 |
| | VBM − 2 | CBM + 1 | (0.4, 0, 0) | 3.33 | 0.29 |
| | VBM − 3 | CBM | (0.4, 0, −0.5) | 3.34 | 0.29 |
| | VBM − 3 | CBM | (−0.4, 0, 0.5) | 3.34 | 0.26 |
| | VBM − 2 | CBM + 1 | (−0.4, 0, 0) | 3.33 | 0.26 |
| $E_{CT}^J = 2.89$ | VBM | CBM + 1 | (−0.2, 0, 0.5) | 3.29 | 0.47 |
| | VBM | CBM + 1 | (−0.2, 0, 0) | 3.28 | 0.41 |
| | VBM | CBM + 1 | (0.2, 0, −0.5) | 3.29 | 0.40 |
| | VBM − 1 | CBM | (−0.2, 0, 0) | 3.31 | 0.37 |
| | VBM | CBM + 1 | (0.2, 0, 0) | 3.28 | 0.36 |
| | VBM − 1 | CBM | (0.2, 0, 0) | 3.31 | 0.32 |
| | VBM − 1 | CBM | (−0.2, 0, 0.5) | 3.29 | 0.26 |
| | VBM − 1 | CBM | (−0.4, 0, 0.5) | 3.23 | 0.24 |
| | VBM | CBM + 1 | (−0.4, 0, 0) | 3.22 | 0.23 |
| | VBM − 1 | CBM | (0.2, 0, −0.5) | 3.29 | 0.23 |

[a]Only weights larger than 2% are reported.

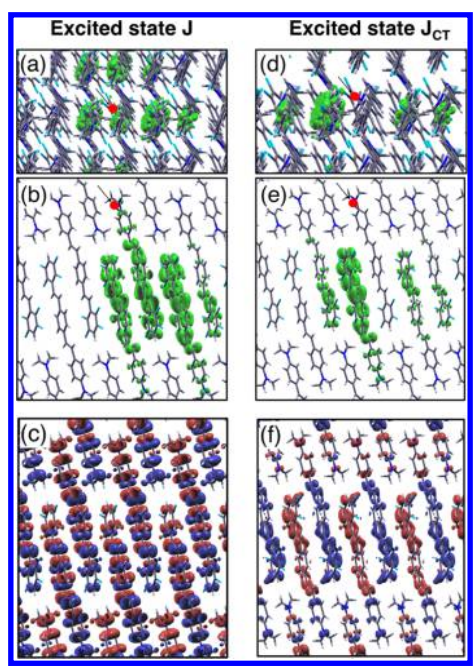

**Figure 5.** (a, b, d, e) Exciton probability density with fixed hole position (red dot), defined as $|\Psi_\lambda(r_e, \bar{r}_h)|^2 = |\langle \lambda | \Psi^\dagger(r_e) \Psi(\bar{r}_h)|0\rangle|^2 = |\sum_{cvk} A_{cvk}^\lambda \varphi_{ck}(r_e) \varphi_{vk}^*(\bar{r}_h)|^2$, where $\Psi_\lambda(r_e, \bar{r}_h)$ is the hole (electron) position, $A_{cvk}^\lambda$ is the BSE coefficient for the excited states $\lambda = J, J_{CT}$ of the J-aggregate, and $\varphi_{vk}(\phi_{ck})$ is the occupied (unoccupied) KS electronic state with wavevector **k** in the BZ; views from the $a$–$b$ plane (a, d) and from $a$–$c$ plane (b, e). The exciton plot gives the probability to find the electron at position $r_e$ with the hole fixed at $\bar{r}_h$. (c, f) Transition density, defined as $\rho_\lambda(r) = \langle \lambda | \Psi^\dagger(r) \Psi(r)|0\rangle = \sum_{cvk} A_{cvk}^\lambda \varphi_{ck}(r) \varphi_{vk}^*(r)$, for the excited states $\lambda = J, J_{CT}$ of the J-aggregate; views from the $a$–$c$ plane. The transition density provides information about the charge spatial displacement associated with the specific excited state $\lambda$. Isosurfaces of the exciton wave functions (transition densities) fixed at 10% of their maximum value.

organic crystals formed by nonpolar molecules.[16,17,79,80] On the other hand, the intramolecular CT nature of the J excitation is inherited from the polarized character of contributing electronic states, in analogy with the MOs of the push−pull molecules constituting the J-aggregate. The transition density of J shown in Figure 5c offers complementary information to the exciton wave function in terms of the spatial distribution of the excitation and the orientation of the molecular transition dipole moments. Since they are coherently aligned and in phase with respect to each other, the excitation J is associated with an induced charge density mainly displaced along the long crystal axis within the $a$–$c$ plane (see Figure 1b). It is also worth noting that, by visual inspection, the transition density shown in Figure 5c does not completely sum to zero within a single monomer unit, as the positive (red) and negative (blue) charge blocks are not present in equal amount. This suggests that there could be a partial intermolecular CT mechanism that has been highlighted above in the analysis of the exciton wave function of Figure 1b. This result has similarities with that obtained for pentacene molecular crystals,[16,19] where exciton delocalization on neighboring molecules also appears. However, as opposed to pentacene, the J-aggregate considered here is composed of polar push−pull molecules: the intrinsic dipole strongly polarizes the frontier MOs and thus reduces the spatial overlap between electron and hole.

The very weak excitation JCT, at slightly lower energy compared to J (see Figure 4b), has a different nature. As shown in Figure 5d−f by the exciton wave function and the transition density, this excitation has a pure intermolecular CT character, with the electron and the hole localized on different neighboring molecules. The slightly larger binding energy of $J_{CT}$ ($E_b = 0.32$ eV) compared to J ($E_b = 0.21$ eV) is consistent with its enhanced spatial localization (Figure 5d,e), while the reduced wave-function overlap between the hole and the electron components is consistent with the very weak oscillator strength. Such a CT excitation is associated with lower electron−hole recombination rates and enhanced electron− hole dissociation probability compared to excitons with more pronounced intramolecular character.[81] The transition density associated with $J_{CT}$ (Figure 5c) confirms and complements this picture: neighboring monomers are almost completely depleted of positive and negative charges, respectively, meaning that intermolecular CT is the dominant mechanism here. It is worth mentioning that the character and the microscopic features of excited states, such as $J_{CT}$, cannot be captured by simple models based on transition dipoles coupling (e.g., the Kasha's model[11]) but need an advanced first-principles description as provided in this work.

The intra- and intermolecular character of singlet excitations in organic crystals is determined by a competition between the e–h exchange interaction, which is responsible for the LFE, and the screened e–h attraction.[19] While the exchange interaction is quite sensitive to the spatial overlap between occupied and unoccupied electronic states involved in the transition, the direct e–h attraction can be nonvanishing even upon a negligible spatial overlap between occupied and unoccupied states.[19] Here, due to the push–pull character of the constituting dyes, the gap states of the J-aggregate are quite polarized and hence only slightly overlap (see Figures 2a and 3), as opposed, for instance, to pentacene.[19] Hence, we should expect a reduced influence from the LFE to the low-lying excited states of the J-aggregate. To quantify the contribution of the LFE on a given excited state $|\lambda\rangle$, we use the e–h exchange energy $E_x^\lambda$ and its ratio with respect to the total excitation energy $\tilde{E}_x^\lambda = E_x^\lambda/E^\lambda$. In the isolated monomer, $E_x^M = 0.76$ eV ($\tilde{E}_x^M = 0.22$), while in the J-aggregate, $E_x^J = 0.07$ eV ($\tilde{E}_x^J = 0.023$) and $E_x^{J_{CT}} = 0.01$ eV ($\tilde{E}_x^{J_{CT}} = 0.003$). From these values, it is apparent that LFE contribute ~20% to the first excitation energy of the isolated monomer while they are almost negligible in the J-aggregate (2% for J and 0.3% for $J_{CT}$). The predominant intermolecular CT character of $J_{CT}$ is related to a weaker e–h exchange interaction with respect to J, where in the latter, the spatial overlap between the electron and the hole is larger (see also Figure 5a,b). The intermolecular CT that characterizes both J and $J_{CT}$ is favored by the close molecular packing[82–84] of the aggregate, which enhances the electron delocalization between neighboring molecules as generally observed in optical excitations of organic crystals.[17,50,53,54,85,86] The reduction of LFE in the J-aggregate compared to the isolated monomer is due to the more homogeneous electron distribution in the crystal.

## CONCLUSIONS

To summarize, in the framework of MBPT, we have investigated the electronic and optical properties of a J-aggregate formed by the push–pull organic dye $C_{24}H_{19}F_4N$, specifically addressing the interplay between intra- and intermolecular interactions. We have found that the intense J-band dominating the absorption onset is formed by a number of excitations stemming from transitions between the highest occupied and lowest unoccupied bands. The most intense of these excitations exhibits a combination of inter- and intramolecular charge transfers, resulting from the competing effects of dense molecular packing and the push–pull nature of the constituting molecules. The other excitations within the J-band have very weak intensity. Among them, $J_{CT}$ has pronounced intermolecular charge-transfer character. Being at lower energy compared to the most intense excitation in the J-band, this state is expected to play a relevant role in the emission properties of the J-aggregate.

Our analysis demonstrates that the complex mechanisms ruling the optical properties of organic crystalline aggregates cannot be unveiled based solely on simple models, but require a high level of theory that is able to quantitatively address all of the facets of the problem. Many-body perturbation theory is capable of thoroughly capturing the collective effects and providing a robust insight into the excitations of the system. As such, our results offer unprecedented insight into the nature of the excitations of J-aggregates formed by push–pull chromophores and contribute to the further understanding of these materials that are relevant for optoelectronic applications.

## ASSOCIATED CONTENT

**Ⓢ Supporting Information**

The Supporting Information is available free of charge on the ACS Publications website at DOI: 10.1021/acs.jpcc.8b11709.

Electronic and optical properties of the isolated push–pull monomer and the J-aggregate (PDF)

## AUTHOR INFORMATION

**Corresponding Authors**
*E-mail: caterina.cocchi@physik.hu-berlin.de (C.C.).
*E-mail: daniele.varsano@nano.cnr.it (D.V.).

**ORCID**
Caterina Cocchi: 0000-0002-9243-9461
Arrigo Calzolari: 0000-0002-0244-7717
Daniele Varsano: 0000-0001-7675-7374
Stefano Corni: 0000-0001-6707-108X

**Notes**
The authors declare no competing financial interest.

## ACKNOWLEDGMENTS

This work was partially funded by the European Union under the ERC grant TAME Plasmons (ERC-CoG-681285), by the Deutsche Forschungsgemeinschaft (DFG, German Research Foundation)—Projektnummer 182087777—SFB 951 and HE 5866/2-1, and by the EU Centre of Excellence "MaX - Materials Design at the Exascale" (Horizon 2020 EINFRA-5, Grant No. 676598). M.G. acknowledges support from the German Academic Exchange Service (DAAD) and HPC-EUROPA3 (INFRAIA-2016-1-730897), with the support of the EC Research Innovation Action under the H2020 Programme. Computational resources were partly provided by PRACE on the Marconi machine at CINECA and by the High-Performance Computing Center Stuttgart (HLRS).